\documentclass[aps,prl,floatfix,twocolumn,preprintnumbers,nofootinbib,reprint,groupedaddress,amssymb,superscriptaddress]{revtex4-1}

\usepackage{amsmath}
\usepackage{amsfonts}
\usepackage{amssymb}
\usepackage{graphicx, rotating}
\usepackage{epstopdf}
\usepackage{epsfig}
\usepackage{latexsym}
\usepackage{mathtools}
\usepackage{graphicx}
\usepackage{color}
\usepackage[dvipsnames]{xcolor}
\usepackage{amsmath,amssymb}
\usepackage{multirow}
\usepackage{array,makecell}
\usepackage[utf8]{inputenc}
\usepackage{mathrsfs}
\usepackage{bbold}

\usepackage{slashed}
\usepackage{hyperref}
\hypersetup{colorlinks, citecolor=bluscuro, linkcolor=bluscuro, urlcolor=bluscuro}
\definecolor{rossos}{cmyk}{0,1,1,0.55}
\definecolor{bluscuro}{rgb}{0.15, 0.2, .85}
\definecolor{bluchiaro}{cmyk}{1,.3,0.,0.1}
\definecolor{verdescuro}{rgb}{0.3,0.8,0.3}

\newcommand{\be}{\begin{equation}}
\newcommand{\ee}{\end{equation}}          
\newcommand{\bea}{\begin{eqnarray}}
\newcommand{\eea}{\end{eqnarray}}
\newcommand{\bc}{\begin{center}}
\newcommand{\ec}{\end{center}}

\def\cale{{\mathcal{E}}}

\def\tr{{\rm tr\,}}

\newcommand{\ab}[1]{\langle #1 \rangle}

\newcommand{\ra}[1]{| #1 \rangle }

\begin{document}

\title{Flow between Extremal One-Point Energy Correlators in QCD}

\preprint{CERN-TH-2025-175}

\author{Marc Riembau}
\affiliation{Theoretical Physics Department, CERN, 1211 Geneva 23, Switzerland}
\affiliation{Theoretical Particle Physics Laboratory, Institute of Physics, EPFL, Lausanne, Switzerland}

\author{Minho Son}
\affiliation{Department of Physics, Korea Advanced Institute of Science and Technology,
	291 Daehak-ro, Yuseong-gu, Daejeon 34141, Republic of Korea}

\begin{abstract}
\noindent 
{
The energy density generated by a vector current is characterized by a single parameter $a_{\mathcal{E}}$ bounded by unitarity to $-1/2 \leq a_{\mathcal{E}} \leq 1$, with extremal values saturated by free theories of different matter content. Through confinement, QCD transmutes fermionic matter into scalars, revealing a nontrivial flow between extremal correlators. 
We reconstruct this flow using perturbative QCD and chiral perturbation theory.
The observable is accessible with currently available experimental data.
}

\end{abstract}

\maketitle

\medskip

%%%%%%%%%%%%%%%%%%%%%
%%%%%%%%%%%%%%%%%%%%%

On its 50th anniversary \cite{Gross:2022hyw}, quantum chromodynamics (QCD) remains a vibrant and rich field, a whetstone for theoretical physicists to sharpen their understanding of quantum field theory and of nature. A defining feature of QCD is confinement, which dramatically changes dynamics as it flows from the ultraviolet (UV) to the infrared (IR). Observables appropriate in one regime often lose their utility in the other: the $S$ matrix efficiently characterizes two-pion scattering, but quickly becomes unfeasible once multiparticle thresholds open, while jet rates are well defined at high energies but not at low.

Event shapes provide a class of observables that can be studied across scales. Indeed, it was through their study that a first understanding of gluon emission emerged \cite{Ellis:1976uc,PhysRevLett.39.1587,PhysRevLett.39.1237,PhysRevLett.41.1581}. A particularly relevant set of event shapes are energy correlators \cite{PhysRevD.17.2298,PhysRevLett.41.1585}, which possess rich theoretical and phenomenological properties, recently reviewed in \cite{Moult:2025nhu}.  
While much of the phenomenological studies have been devoted to multipoint correlators, the simplest case, the one-point energy correlator, remains largely unexplored.

The angular dependence of the one-point energy correlator of a vector current is fully characterized by a single parameter $a_\cale$ 
\be
\ab{\cale_n}\,=\, \frac{\ab{\cale}}{4\pi}\left[ 1 + a_\cale\left( \frac32\sin^2\theta-1 \right)\right]~,
\label{eq:onepointEexplicit}
\ee
with $\ab{\cale}$ the total energy. 
The parameter $a_\cale$ is constrained by energy positivity to lie within $-1/2 \leq a_\cale \leq 1$ \cite{Hofman:2008ar}. 
While free theories with different matter content saturate the different extremal values of $a_\cale$ \cite{Zhiboedov:2013opa}, strongly coupled theories that transmute their degrees of freedom may transition from one boundary to the other. 
In this Letter, we observe that QCD provides a striking realization of this phenomenon, flowing between the two extremal energy correlators as the theory converts fermionic degrees of freedom in the UV into bosonic ones in the IR.

\begin{figure*}
	\centering
	\includegraphics[width=\linewidth]{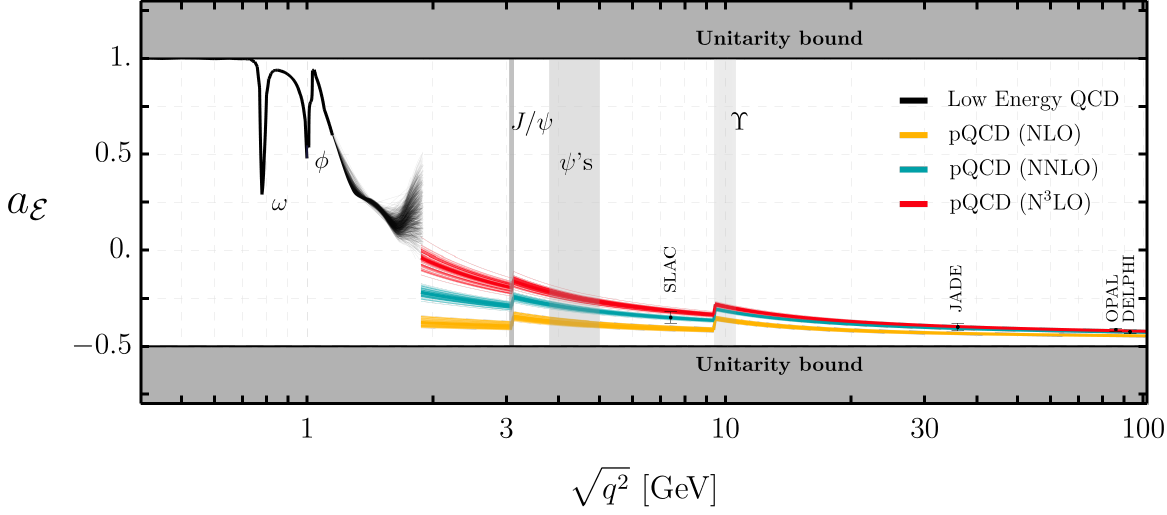}	\caption{\small QCD flow for $a_\cale$ between the fermionic (UV, $a_\cale\to -1/2$) and bosonic (IR, $a_\cale\to 1$) extremal correlators. 
		The gray region is forbidden by unitarity. 
		The black solid line shows the predictions from $\chi$PT and data-driven methods. Spread of black lines shows the uncertainty associated with missing hadronic channels. 
		The colored lines represent the N${}^k$LO predictions in perturbative QCD (pQCD), with line spread given by the uncertainty associated with $m_c$, $m_b$, $\alpha_s(m_Z^2)$ and the renormalization scale $\mu$. We indicate the charmonium and bottomonium regions.  Experimental measurements of Table~\ref{tab:aE_measurements} are shown.
	}
	\label{fig:onepoint}
\end{figure*}

The goal of this Letter is to determine the global evolution of $a_\cale$ from the electroweak scale all the way down to the pion threshold by combining the predictions from perturbative QCD with the ones of chiral perturbation theory. 
Our main result is summarized in Figure~\ref{fig:onepoint}. The remainder of the Letter is devoted to outlining the theoretical framework and key ingredients that lead to it.

%%%%%%%%%%%%%%%%%%%%%%%%%%%%%%%%
%%%%%%%%%%%%%%%%%%%%%%%%%%%%%%%%
\medskip
\textbf{\textit{Flowing one-point energy correlator.}}\,\, 
The one-point energy correlator measured along $\vec{n}$ on a state sourced by the electromagnetic vector current $J^\mu$ is defined from the three-point correlator between the currents and the energy flow operator $ \mathcal{E}_n$
\be
H^{\mu\nu}_\cale\,=\,\int d^4 x e^{i q \cdot x} \langle J^{\mu\dagger}(x) \mathcal{E}_n J^\nu(0)\rangle~.
\label{eq:3ptcorrelator}
\ee
Dotting it to a polarization basis along the current spin axis allows to define the density matrix $
\langle \cale_n \rangle_{h^\prime h} \,=\, \epsilon^{*\mu}_{h^\prime} H^{\mu\nu}_\cale \epsilon^\nu_h$. 
Given that the correlator only depends on the total injected momentum $q^\mu$ and the detector direction $n^\mu$, in the rest frame with $q^\mu=(\sqrt{q^2},\vec{0})$ and $n^\mu = (1,\vec{n})$, the density matrix is fully determined in terms of a single parameter $a_\cale$
\be
\langle \cale_n \rangle_{h^\prime h} = \frac{\ab{\cale}}{4\pi}\left[\delta_{h^\prime h} + a_\cale \left( 3(n\cdot \epsilon^*_{h^\prime})(n\cdot \epsilon_h)-\delta_{h^\prime h} \right) \right]~,
\label{eq:densitymatrix}
\ee
where $\ab{\cale}=\sqrt{q^2}$ is the total energy injected in the current and $a_\cale$ measures the relative weight between the inclusive and the spinning part of the correlator.

For the unpolarized case of $e^+e^-\to \gamma^*\to \text{hadrons}$, the beam produces the trace of transverse polarizations  $\frac{1}{2}\tr\langle \cale_n \rangle_{h^\prime h}$, which implies $\delta_{h^\prime h}\to 1$ and $(n\cdot \epsilon^*_{h^\prime})(n\cdot \epsilon_h)\to \frac12\sin^2\theta$ in Eq.~\ref{eq:densitymatrix}, where $\theta$ is the angle between spin axis of the current, in this case coinciding with the beam axis, and the detector direction. 
The explicit form of the one-point energy correlator is given in Eq.~\ref{eq:onepointEexplicit}. 

The two independent Lorentz scalars that can be built out of the three point correlator in Eq.~\ref{eq:3ptcorrelator} are $-\eta_{\mu\nu} H^{\mu\nu}_\cale$ and $n_\mu n_\nu H^{\mu\nu}_\cale$. They determine $a_\cale$ as
\be
a_\cale\,=\, -\frac12\left(1-3\,\frac{n_\mu n_\nu H^{\mu\nu}_\cale}{-\eta_{\mu\nu} H^{\mu\nu}_\cale}\right)~,
\label{eq:aEratio}
\ee
depending on the total energy $q^2$. 
The unitarity and positivity of the energy imply that the contractions of the correlator obey the constraints $n_\mu n_\nu H^{\mu\nu}_\cale \geq 0$ and $-\eta_{\mu\nu} H^{\mu\nu}_\cale-n_\mu n_\nu H^{\mu\nu}_\cale\geq 0$, which imply the bounds $
-1/2 \,\leq\, a_\cale\,\leq\, 1\,$,
for the $a_\cale$ parameter controlling the shape of the energy density \cite{Hofman:2008ar}. 
The $a_\cale$ depends on the average angular momentum of the generated state along the $\vec{n}$ direction. Upper and lower bounds are given by states with $m=0$ and $|m|=1$ spin along $\vec{n}$, respectively. 
Examples include a current of minimally coupled fermions ($a_\cale=-1/2$), scalars ($a_\cale=1$), or a Wess-Zumino-Witten (WZW) term $\epsilon^{\mu\nu\rho\sigma}\partial_\nu\pi^a\partial_\rho \pi^b\partial_\sigma\pi^c$ ($a_\cale=-1/2$). 

In  the following, we present the calculation of the evolution of $a_\cale$ as a function of the momentum $q^2$, tracking its flow from the UV to the IR.

\medskip

\textit{---Perturbative QCD.} \quad 
We can compute the correlator in Eq.~\ref{eq:3ptcorrelator} in perturbation theory by noting that the energy operator $\cale_n$ acts on multiparticle states $\ra{\alpha}$ as $\cale_n\ra{\alpha} = \sum_i E_i \delta^{(2)}(\Omega_n-\Omega_i)\ra{\alpha}$. This leads a perturbative representation of the correlator as squared matrix elements with a phase space weighted by the energy and a sum over all states with particles in the direction $\vec{n}$. 

The perturbative determinations of $-\eta_{\mu\nu} H^{\mu\nu}_\cale$ and $n_\mu n_\nu H^{\mu\nu}_\cale$ can actually be derived from the computations of the longitudinal and transverse fragmentation functions in $e^+e^-$ annihilation \cite{Nason:1993xx}. 
The semi-inclusive process $e^+e^-\to hX$, where $h$ represents a given hadron species and $X$ is the rest of the state, is described by separating the cross section into a transverse ($T$), longitudinal ($L$), and asymmetric ($A$) part, each one having an inclusive cross section of the form
\be
\frac{d\sigma^h_P}{dx} = \sum_i\int_x^1\frac{dz}{z} C_{P,i}(z,q^2,\mu) D_i^h(x/z,\mu)~,
\ee
where $P=T,\, L,\, A$ and $i$ runs over the types of partons. $C_{P,i}$ is calculable in perturbation theory and $D_i^h$ is the so-called fragmentation function for $h$ from $i$ partons. In the energy correlator, one is fully inclusive over $h$ and the type of parton $i$, which implies sensitivity only to the hard coefficients $C_{P,i}$. Indeed, using the fact that $\sum_h\int_0^1 dz z D^h_i(z,\mu)=1$ due to the energy conservation, this allows us to represent tensor structures of the energy correlator at high energies in terms of integrals of $C_{P,i}$,
\bea\nonumber
	-\eta_{\mu\nu} H^{\mu\nu}_\cale &=& \frac12 \sum_i \int_0^1 dx\,x\left[ C_{T,i}(x,q^2) + C_{L,i}(x,q^2) \right]~,
	\\
	n_\mu n_\nu H^{\mu\nu}_\cale &=& \frac12 \sum_i \int_0^1 dx\,x\, C_{L,i}(x,q^2)~.
	\label{eq:etannintermsofC}
\eea
The factor $x$ in the integrand corresponds to the energy weight, and it annihilates the soft part of the coefficient functions. The collinear divergences given by $1/\epsilon$ terms in $C_{P,i}$ are proportional to the parton splitting functions. They vanish after being inclusive over the partons $i$, implied by the collinear safety of the observable.  
The asymmetric part does not enter in the one-point energy correlator, but it does enter in the parity-odd terms of the one-point charge correlator of an electroweak current \cite{Riembau:2024tom}.

It is interesting to realize that at Born level $n_\mu n_\nu H^{\mu\nu}_\cale = 0$, and therefore, the boundary $a_\cale = -1/2$ is saturated. This is equivalent to the Callan-Gross relation, which implies that at high energies the electromagnetic current is made out of only spin-1/2 fields \cite{PhysRevLett.22.156}. 
At next-to-leading order (NLO), the correction is the same for both tensor structures, given by $\frac32 (\alpha_s/2\pi) C_F\,$ 
\cite{PhysRevD.17.2298,PhysRevLett.41.1585}. 
At NNLO, the corrections for $C_{P,i}$ were obtained in \cite{Rijken:1996ns,Rijken:1996vr,Mitov:2006wy,Blumlein:2006rr}. 
The $\alpha_s^3$ calculation of the hard coefficient functions $C_{P,i}$ has been recently achieved in \cite{He:2025hin}. By taking the latter results and performing the integrals, with the help of the packages \verb*|HPL| and \verb*|PLT| \cite{Maitre:2005uu,Duhr:2019tlz}, we get the $\alpha_s^3$ expressions for $-\eta_{\mu\nu} H^{\mu\nu}_\cale$ and $n_\mu n_\nu H^{\mu\nu}_\cale$ reported in the Appendix. 
As a validation, the expression for $-\eta_{\mu\nu} H^{\mu\nu}_\cale$ matches the numerical expression for the $\mathcal{O}(\alpha_s^3)$ term for the hadronic $R$ ratio in \cite{Baikov:2012zn}. Expanding the two Lorentz contractions in Eq.~\ref{eq:aEratio} up to order $\alpha_s^3$, it leads to the N${}^3$LO expression for the parameter $a_\cale$ in perturbative QCD for massless quarks

\begin{widetext}
	\begin{equation}
		\begin{split}
			a_\cale\,\,=\,\,&
			-\frac12
			\,+\,\frac{\alpha_s}{\pi}\frac{9C_F}{8}\,+\,\left(\frac{\alpha_s}{\pi}\right)^2\left[ -C_F^2\frac{99}{64} + C_F C_A\left( \frac{2023}{320} -\frac{9}{20}\zeta_3 \right) -C_F n_F \frac{37}{32} \right]
			\\%[2pt]
			&+ \left(\frac{\alpha_s}{\pi}\right)^3\bigg[ \,\,\,C_F^3 \left( -\frac{3}{1280}-\frac{537}{40}\zeta_3+\frac{69}{4}\zeta_5 \right)
			+C_F^2 C_A \left( -\frac{14737}{960}-\frac{165}{16}\zeta_2+\frac{3803}{160}\zeta_3+\frac{605}{32}\zeta_4-\frac{231}{8}\zeta_5  \right)
			\\%[2pt]
			&\hspace{1.8cm} +C_F C_A^2 \left( \frac{2599751}{57600}+\frac{319}{64}\zeta_2-\frac{307}{25}\zeta_3-\frac{209}{20}\zeta_4+\frac{87}{16}\zeta_5  \right)
			\\%[2pt]
			&\hspace{1.8cm} +C_F^2 n_f \left( \frac{6163}{3840}+\frac{15}{8}\zeta_2-\frac{71}{40}\zeta_3-\frac{55}{16}\zeta_4+3 \zeta_5  \right) +C_F n_f^2 \left( \frac{461}{360}-\frac{1}{16}\zeta_2+\frac{3}{40}\zeta_3 \right)
			\\%[2pt]
			&\hspace{1.8cm} +C_F C_A n_f \left( -\frac{3622}{225}-\frac{9}{16}\zeta_2+\frac{259}{400}\zeta_3+\frac{19}{10}\zeta_4+\frac{3}{8}\zeta_5  \right)
			+\frac{d^{abc}_Fd^{abc}_F}{N_c} \left( \frac{69}{40}+\frac{42}{5}\zeta_3-12\zeta_5  \right)\,\,\, \bigg]~.
			\label{eq:aEresultPT}
		\end{split}
	\end{equation}
\end{widetext}

For finite fermion masses, both tensor structures are nonvanishing at Born level, $-\eta_{\mu\nu} H^{\mu\nu}_\cale=1+n_\mu n_\nu H^{\mu\nu}_\cale=1+\frac12 (4m_f^2/q^2)$. 
Near threshold $a_\cale\to 0$ and the vector current behaves like a scalar operator.  
Compared with loop corrections of order $\alpha_s/\pi$, this effect from the mass is relevant for $q^2\lesssim 4m_f^2 (\pi/\alpha_s)$, and we shall see that it is sizable for $b$ quarks even away from threshold.

In QCD with $n_f=5$, all numerical coefficients of $\alpha_s^k$ in Eq.~\ref{eq:aEresultPT} are all positive. At $q^2=m_Z^2$ and using $\alpha_s(m_Z^2)=0.118$, 
one gets $a_\cale=-0.4437$, $a_\cale=-0.4258$ and $a_\cale=-0.42$ at NLO, NNLO and N${}^3$LO, respectively. 

Even though there is no experimental measurement explicitly dedicated to the one-point energy correlator, we observe that the expressions in Eq.~\ref{eq:etannintermsofC} directly relate $a_\cale$ to existing measured observables through
\be
a_\cale = -\frac12 + \frac32 \frac{\sigma_L}{\sigma_{\text{tot}}}~,
\ee
where $\sigma_L$ is the longitudinal cross section in semi-inclusive $e^+e^-\to hX$. In this representation, the positivity constraints are simply $\sigma_{T,L}\geq 0$. The longitudinal ratio $\sigma_L/\sigma_{\rm tot}$ has been measured at LEP by both OPAL and DELPHI Collaborations in \cite{OPAL:1995xgn,DELPHI:1997oih} at the $Z$ pole. The JADE Collaboration, at PETRA, also reported a measurement at $\sqrt{q^2}=36.6\,\text{GeV}$ \cite{Blumenstengel:2001yi}. 
The data at SLAC at $7.4\,\text{GeV}$ in \cite{Schwitters:1975dm,Feldman:1977nj} may also be interpreted in terms of $\sigma_L/\sigma_{\rm tot}$, although with statistical errors only, as done in \cite{Blumenstengel:2001yi}. The interpretation of these measurements in terms of $a_\cale$ is summarized in Table~\ref{tab:aE_measurements}.

The LEP measurements agree well with the N${}^3$LO results. The JADE value is compatible with the N${}^3$LO result of $a_\cale=-0.396$ at $\sqrt{q^2}=36.6\,\text{GeV}$, obtained after including the $b$-quark mass effect. The SLAC measurement does not contain systematic uncertainties, but seems to validate the perturbative approach. Theoretical and parametric uncertainties will be discussed with the results in Fig.~\ref{fig:onepoint}. 
It should be emphasized that, despite their equivalence, dedicated experimental measurements of $a_\cale$ through the energy correlator should be simpler and cleaner than $\sigma_L/\sigma_{\rm tot}$ measurements.

\begin{table}[t]
	\centering
	\renewcommand{\arraystretch}{1.25} 
	\begin{tabular*}{\columnwidth}{@{\extracolsep{\fill}} l c c c}
		\hline\hline
		Experiment & $\sqrt{s}$ [GeV] & $a_\mathcal{E}$ & Ref. \\
		\hline
DELPHI      & 91.2    & $-0.423 \pm 0.011$ & \cite{DELPHI:1997oih} \\
		OPAL        & 91.2    & $-0.415 \pm 0.008$ & \cite{OPAL:1995xgn} \\
		JADE        & 36.6  & $-0.40 \pm 0.02$   & \cite{Blumenstengel:2001yi} \\
		SLAC        & 7.4   & $-0.35 \pm 0.03$   & \cite{Feldman:1977nj,Blumenstengel:2001yi} \\
		\hline\hline
	\end{tabular*}
	\caption{Indirect determinations of $a_\mathcal{E}$ from 
		$\sigma_L/\sigma_{\rm tot}$ measurements in $e^+e^-$ data at different center of mass energies.}
	\label{tab:aE_measurements}
\end{table}

\medskip

\textit{---Low energy QCD.} \quad 
At low energies, due to the mass gap, processes involve a finite small number of particles. 
QCD can be studied using chiral perturbation theory ($\chi$PT) as well as through dispersive methods. For a recent review of dynamics below $\sqrt{q^2} \lesssim 2~\text{GeV}$, see \cite{Fang:2021wes}.

The contributions to $a_\cale$ from two and three pseudoscalar states, which dominate the cross section at low energy, are fixed by symmetries. 
Two-pseudoscalar states saturate the upper bound $a_\cale=1$. This can be explicitly seen since the hadronic current is fixed to have the form $J^\mu = (p_1^\mu-p_2^\mu) F^V_P(q^2)$ where $p_{1,2}^\mu$ are the pseudoscalar momenta, and $F^V_P(q^2)$ is a form factor that defines the meson's charge via $F^V_P(0)=Q_P$. The matrix element leads to the relation $-\eta_{\mu\nu}H^{\mu\nu}_\cale = n_\mu n_\nu H^{\mu\nu}_\cale$, saturating the upper bound of $a_\cale$ independently of the form factor and pseudoscalar mass. 
It should be noted that $\alpha_\text{em}$ corrections to this extremal value, leading to final state radiation, can only decrease $a_\cale$ due to unitarity.

For states with three pseudoscalars, the structure of the hadronic current is determined by parity considerations, given by $J_\mu= \epsilon_{\mu\nu\alpha\beta}p_1^\nu p_2^\alpha p_3^\beta \mathcal{F}(s_1,s_2,s_3)
$
~\cite{Hoferichter:2018kwz}, 
with $\mathcal{F}(s_1,s_2,s_3)$ being some function of $s_i=(q-p_i)^2$ fixed by the WZW term in the $s_i\to 0$ limit \cite{Adler:1971nq,Aviv:1971hq}. 
Independently of the form factor $\mathcal{F}(s_i)$, the tensor structure implies that the projection of the correlator on $n^\mu n^\nu$ vanishes, $n_\mu n_\nu H^{\mu\nu}=0$, thus, saturating the lower bound $a_\mathcal{E}=-1/2$. 
The vector-pseudoscalar channels, like $\omega\pi$ and $\omega\eta$, are also fixed by symmetry and saturate the lower bound $a_\cale=-1/2$, 
as it can be explicitly seen from the  $\epsilon^{\mu\nu\lambda\delta}\partial_\mu\omega_\nu\partial_\lambda \rho^i_\delta\pi^i$ vertex \cite{Fujiwara:1984mp}. It is worth noticing how all currents leading to $a_\cale=-1/2$ are induced by the WZW term.

The two and three pseudoscalar channels saturate the total rate up to $1\,\text{GeV}$, where the four-pion final states quickly start to dominate the cross section. 
While near the  $4\pi$ threshold $\chi$PT predictions for its production are still in agreement with data \cite{Ecker:2002cw}, in the energy range $\sqrt{q^2}\lesssim 2\,\text{GeV}$ of interest to determine $a_\cale$, the derivative expansion of $\chi$PT breaks down. 
The phenomenological model proposed in \cite{Czyz:2008kw}, based on \cite{Czyz:2000wh,Druzhinin:2007cs} and implemented in \texttt{PHOKHARA} \cite{Rodrigo:2001kf,Campanario:2019mjh}, allows to fit the $e^+e^-$ data up to $\sqrt{q^2}\simeq 3\,\text{GeV}$. Using this numerical approach, we estimate $a_\cale$ for the $4\pi$ channels. The results are shown in Fig.~\ref{fig:onepoint:sigma:aEeach} of the Appendix.

The aforementioned channels dominate the cross section up to $\sqrt{q^2}\simeq 1.4\,\text{GeV}$, above which many other channels open. 
We take into account the contributions of these channels by extracting the exclusive contributions to the total hadronic $R$ ratio between $1.12\,\text{GeV}$ and $1.937\,\text{GeV}$ from the analysis in \cite{Keshavarzi:2018mgv}. 
For these channels, notably for $KK\pi\pi$, $5\pi$ and $6\pi$, reliable predictions for $a_\cale$ are currently unavailable. Therefore, they become the main source of uncertainty around $2\,\text{GeV}$. The impact of this uncertainty in the evolution of $a_\cale$ will be detailed in the following discussion of Fig.~\ref{fig:onepoint}. 

Since the extremality of the two and three pseudoscalar channels crucially relies on the precise isolation of the initial state radiation, it provides a nontrivial consistency test of its treatment. This could have an impact on the assessment of their contribution to the hadronic vacuum polarization.

\medskip
\textbf{\textit{Results.}}\,\,
The flow of the $a_\cale$ parameter is presented in Fig.~\ref{fig:onepoint}, combining perturbative QCD at high energies with $\chi$PT and data-driven input at low energies.

At large $q^2$, the perturbative QCD results, obtained using Eq.~\ref{eq:aEratio} with $-\eta_{\mu\nu} H^{\mu\nu}_\cale$ and $n_\mu n_\nu H^{\mu\nu}_\cale$ are shown  at NLO (yellow), at NNLO (teal) and at N${}^3$LO (red). In these, we consider the finite mass effects for the $b$- and $c$-quarks, only at Born level, while light quarks are treated as massless. 
The spread of lines for each result in  Fig.~\ref{fig:onepoint} reflects their uncertainty, estimated by drawing the input parameters from a gaussian distribution around the current experimental ranges: $m_b = 4.18 \pm 0.03,\text{GeV}$, $m_c = 1.27 \pm 0.02,\text{GeV}$, and $\alpha_s(m_Z^2) = 0.118 \pm 0.0016$ \cite{ParticleDataGroup:2024cfk}, and the renormalization scale $\mu$ uniformly over the range $\mu = \sqrt{q^2}\times 2^{\pm 1}$. 
The uncertainty of the NLO result is dominated by the scale variation. 
On the contrary, the N${}^3$LO uncertainty is dominated by the one associated with input parameters, mostly $\alpha_s(m_Z^2)$. 
The grey vertical bands indicate presence of resonances in the charmonium and bottomonium region.

The four-loop calculation of the hadronic vacuum polarization \cite{Baikov:2008jh} shows good agreement with nonresonant experimental data in the energy range $\sqrt{q^2} \in [1.8, 3.7]\,\text{GeV}$ \cite{Davier:2019can}. 
It is natural to expect a similar agreement for $a_\cale$. 
This motivates the tentative extension of the perturbative results for $a_\cale$  in Fig.~\ref{fig:onepoint} up to $\sqrt{q^2}\simeq 2\,\text{GeV}$.  Given the difference between the mean value of the N${}^3$LO and the previous order, it is important to assess whether the $\alpha_s^4$ correction stabilizes. 
It would also be valuable to include $\Lambda_{QCD}/\sqrt{q^2}$ corrections.

The four data points in Fig.~\ref{fig:onepoint} represent the recast of the experimental measurements as summarized in Table~\ref{tab:aE_measurements}. To our best knowledge, this is the first interpretation of data in terms of $a_\cale$.

In the low energy region, the black curve from the pion threshold up to $1.12\,\text{GeV}$ shows the prediction from data-driven methods as obtained with the \texttt{PHOKHARA} Monte Carlo. 
The two-pion states saturate $a_\cale=1$, while the $\omega$ and $\phi$ resonances give sharp features due to the sudden enhancement of the rate for tree-pion states, which saturate the $a_\cale=-1/2$ boundary. 
Above the $\phi$ resonance, the four-pion states start to contribute, rapidly pushing down the value of $a_\cale$.

Between $1.12\,\text{GeV}$ and $1.936\,\text{GeV}$ we use the rates as extracted from data in \cite{Keshavarzi:2018mgv} with their $a_\cale$ predictions. For the channels that we do not have a reliable prediction for $a_\cale$, like the $KK\pi\pi$, $5\pi$, $6\pi$, $\omega\eta\pi$ and others, we consider a conservative flat prior for $a_\cale$ within the unitarity bounds. Drawing multiple samples from this prior leads to the spread of black lines in Fig.~\ref{fig:onepoint}.% 

Around $2\,\text{GeV}$, both high and low energy approaches are at the limits of their validity. Nonetheless, they give compatible predictions for $a_\cale$, giving a coherent global picture for its evolution in QCD.

%%%%%%%%%%%%%%%%%%%%%
%%%%%%%%%%%%%%%%%%%%%
\medskip
\textbf{\textit{Conclusions.}}\,\,
We have presented the first determination of the flow of the one-point energy correlator in QCD, showing its transition between the extremal correlators as it runs from the UV to the IR. This has been done by combining perturbative QCD at high energies and $\chi$PT and data-driven methods at low energies. We observed a simple connection between $a_\cale$ and existing experimental measurements allowing us to extract $a_\cale$ from data for the first time. 
Future improvements on theoretical predictions at both high and low energies should be made in order to resolve the transition region around $2\,\text{GeV}$. 
The precise reconstruction of the $a_\cale$ flow from the wealth of experimental data seems within reach.

The one-point energy correlator, by virtue of its theoretical and experimental simplicity, is a clean and powerful probe of QCD across scales, and a new benchmark for testing the emergence of hadronic degrees of freedom.

%%%%%%%%%%%%%%%%%%%%%
%%%%%%%%%%%%%%%%%%%%%
%\subsection*{Acknowledgments}

\medskip
Acknowledgments---\, 
We thank Gauthier Durieux, Heejoo Kim, Lorenzo Ricci and Francesco Riva for discussions, suggestions and comments on the draft. 
MS was supported by National Research Foundation of Korea under Grant Number RS-2024-00450835.

\bibliographystyle{utphys}
\bibliography{bibs}

%\newpage

\onecolumngrid
\newpage
\appendix

%%%%%%%%%%%%%%%%%%
%\section*{Supplementary Material}
\section*{End Matter}
\label{app:supplemental}
\begin{figure}[h!]
	\centering
	\includegraphics[width=0.33\linewidth]{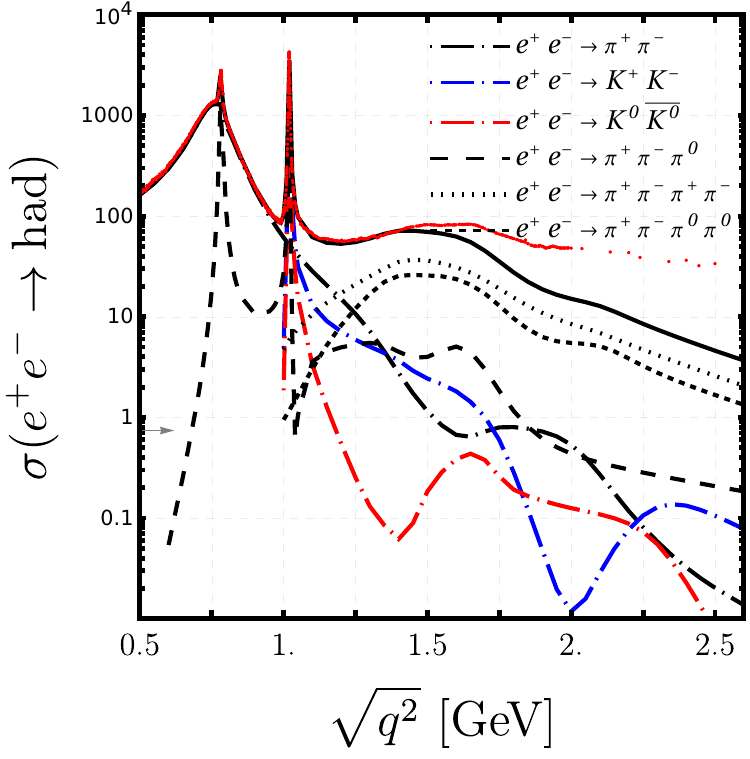}\hspace{2cm}
	\includegraphics[width=0.337\linewidth]{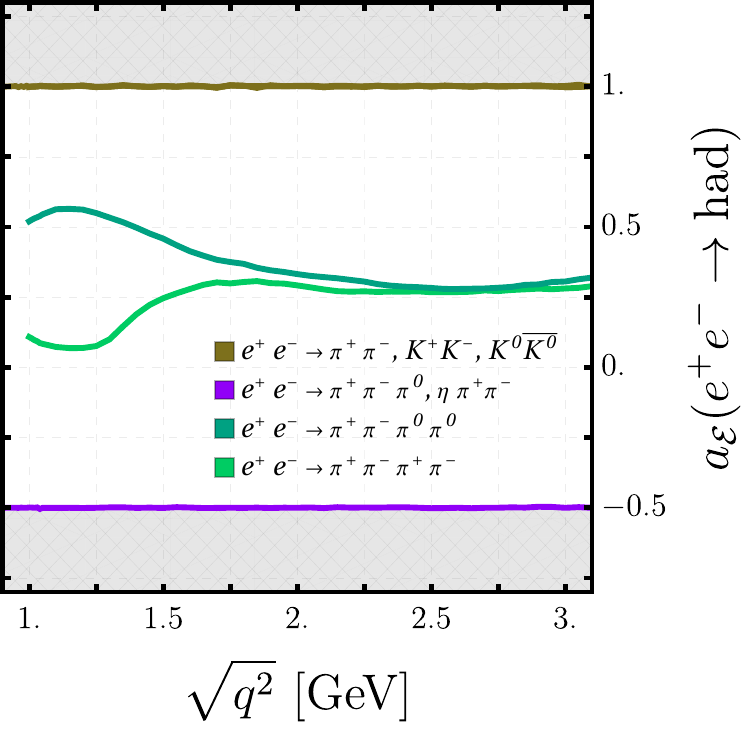}
	\caption{\small Left: Total cross section for $e^+e^-\rightarrow \text{hadrons}$ and decomposition into subprocesses. The solid black line is total sum of six sub-processes by the Monte Carlo simulation with \texttt{PHOKHARA~10.0}. The red dots were extracted from $R$ ratio in PDG data. Difference between both curves is due to other channels becoming important. Right:  evolution of $a_\cale$ of the one-point correlator for individual processes. The pair-production of pseudoscalars, $e^+e^-\rightarrow \pi^+\pi^-$, $K^+K^-$, $K^0 \bar{K^0}$,  all saturate the upper limit of $a_\cale = 1$, while $e^+e^-\to 3\pi$ saturates the lower limit $a_\cale=-1/2$.}
	\label{fig:onepoint:sigma:aEeach}
\end{figure}
\textit{Appendix}---The left panel of Fig.~\ref{fig:onepoint:sigma:aEeach} is the reproduction of the $e^+e^-\to\text{hadrons}$ individual cross sections as reproduced by the \texttt{PHOKHARA} Monte Carlo. The total sum of the channels (black solid curve ) saturates the total cross section from data (red dots) up to $\sim$1.2 GeV. The uncertainty on $a_{\mathcal E}$ up to $\sim$1.2 GeV is dominated by the uncertainty of the ratio of cross section of two-pion vs three-pion production. These, including the $\omega$ and $\phi$ resonant regions, are measured at the percent level~\cite{Belle-II:2024msd}. 
The right panel of Fig.~\ref{fig:onepoint:sigma:aEeach} illustrates the individual contributions to $a_\mathcal{E}$ from each subchannel. For each channel, we fitted the average energy distribution at fixed $q^2$, obtaining $a_\cale$. The number of events are sufficiently large so the statistical uncertainty from the Monte Carlo simulation is negligible. 

\begin{figure}
	\centering
	\includegraphics[width=0.36\linewidth]{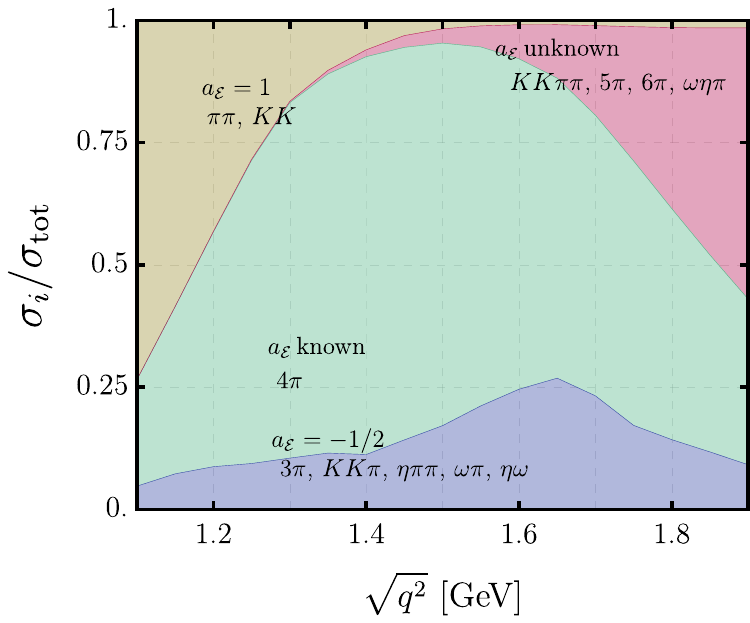}
	\caption{Relative weight of the hadronic channels between $1.12\,\text{GeV}$ and $1.936\,\text{GeV}$. The colors group channels into 4 groups, depending on the value of $a_\cale$ for those.}
	\label{fig:ratioxs_stack}
\end{figure}

In the region between $1.12\,\text{GeV}$ and $1.936\,\text{GeV}$ we extract the relative cross sections from each channel using the analysis of the experimental data done in \cite{Keshavarzi:2018mgv}. In Fig.~\ref{fig:ratioxs_stack}, we show the relative cross sections of the different processes grouped into four categories according to their contributions to $a_\mathcal{E}$. In blue, a group of process saturating the lower bound $a_\cale = -1/2$, given by the two-pseudoscalar channels. In yellow, the channels saturating the upper bound $a_\cale = 1$, given by the three-pseudoscalar channels and the vector-pseudoscalar ones. In green, the channels that have an $a_\cale$ in the bulk, but that we estimated via the numerical simulation, namely the $4\pi$ channels. In red, the channels that have an $a_\cale$ in the bulk, and that we do not have a reliable estimate for it, namely the $KK\pi\pi$, $5\pi$, and $6\pi$ channels.

The expressions for two scalar contractions of the one-point correlator up to order of $\alpha_s^3$, obtained from integrating the results of \cite{He:2025hin}, and setting $\mu_R^2 = \mu_F^2 = q^2$ are given by
\begin{equation}
	\begin{split}
		-\eta_{\mu\nu}H^{\mu\nu} =
		1 + &\left ( \frac{\alpha_s}{2\pi} \right ) \frac{3}{2}C_F
		\\[2pt]
		+ &\left ( \frac{\alpha_s}{2\pi} \right )^2 \left [ C_F^2 \left(-\frac38\right) + C_F C_A \left(\frac{123}{8} - 11\zeta(3)\right) + C_F n_f T_f \left( -\frac{11}{2}+ 4 \zeta(3) \right) \right ]
		\\[2pt]
		+ &\left ( \frac{\alpha_s}{2\pi} \right )^3 \left [  -C_F^3 \frac{69}{16}+C_F^2 C_A \left( -\frac{127}{8}-\frac{143}{2}\zeta_2+110\zeta_5  \right) \right .
		\\[2pt]
		&\hspace{1.6cm}+C_F C_A^2 \left( \frac{90445}{432}-\frac{121}{12}\zeta_2-\frac{2737}{18}\zeta_3-\frac{55}{3}\zeta_5  \right) 
		+C_F^2 n_f \left( -\frac{29}{16}+19\zeta_3-20 \zeta_5  \right)
		\\[2pt]
		&\hspace{1.6cm}
		+C_F n_f^2 \left( \frac{151}{27}-\frac{1}{3}\zeta_2-\frac{38}{9}\zeta_3 \right)
		+C_F C_A n_f \left( -\frac{1940}{27}+\frac{11}{3}\zeta_2+\frac{448}{9}\zeta_3+\frac{10}{3}\zeta_5  \right)
		\\[2pt]
		&\hspace{1.6cm}\left .+\frac{d^{abc}_Fd^{abc}_F}{N_c} \left( \frac{22}{3}-16\zeta_3 \right) \right ]~,
		\label{eq:Heta3}
	\end{split}
\end{equation}
and
\begin{equation}
	\begin{split}
		n_\mu n_\nu H^{\mu\nu} =\quad
		&\left ( \frac{\alpha_s}{2\pi} \right ) \frac{3}{2}C_F
		\\[2pt]
		 + &\left ( \frac{\alpha_s}{2\pi} \right )^2 \left [ C_F^2 \left(-\frac{15}{8}\right) + C_F C_A \left(\frac{2023}{120} - \frac65\zeta(3)\right) + C_F n_f T_f \left( -\frac{37}{6} \right) \right ]
		\\[2pt]
		+ &\left ( \frac{\alpha_s}{2\pi} \right )^3 \left [ C_F^3 \left( -\frac{541}{80}-\frac{358}{5}\zeta_3+92\zeta_5  \right)
		+C_F^2 C_A \left( -\frac{3017}{90}-55\zeta_2+\frac{1627}{15}\zeta_3+\frac{605}{6}\zeta_4-154\zeta_5  \right) \right .
		\\[2pt]
		&\hspace{1.6cm}+C_F C_A^2 \left( \frac{2599751}{10800}+\frac{319}{12}\zeta_2-\frac{4912}{75}\zeta_3-\frac{836}{15}\zeta_4+29\zeta_5  \right)
		\\[2pt]
		&\hspace{1.6cm}+C_F^2 n_f \left( -\frac{137}{720}+10\zeta_2-\frac{97}{15}\zeta_3-\frac{55}{3}\zeta_4+16 \zeta_5  \right)
		+C_F n_f^2 \left( \frac{922}{135}-\frac{1}{3}\zeta_2-\frac{2}{5}\zeta_3 \right)
		\\[2pt]
		&\hspace{1.6cm} +C_F C_A n_f \left( -\frac{57952}{675}-3\zeta_2+\frac{259}{75}\zeta_3+\frac{152}{15}\zeta_4+2\zeta_5  \right)
		\\[2pt]
		%
		%&
		&\hspace{1.6cm}\left .+\frac{d^{abc}_Fd^{abc}_F}{N_c} \left( \frac{46}{5}+\frac{224}{5}\zeta_3 -64\zeta_5 \right) \right ]~.
		\label{eq:Hn3}
	\end{split}
\end{equation}

\end{document}